# What is the right form of the probability distribution of the conductance at the mobility edge?

In a recent letter, Slevin and Ohtsuki [1] reported finite size scaling results for the Anderson metal to insulator transition for the orthogonal and unitarity classes of the single electron tight-binding(TB) model. The average value of the conductance $G=(e^2/h)g$ at the mobility edge, as well as the distribution of the conductance at the critical point, $p_c(g)$, were calculated. Their studies showed that $p_c(g)$ is independent of the system size. It also does not show any dip around g=0, as the $\epsilon$ expansion results [2] suggest. These conclusions were based on numerical results of system sizes of N x N x N, with N=6,8 and 10. We will present new numerical data that indeed shows that $p_c(g)$ has a dip for small g.

We have systematically studied the conductance G of the 3d TB model by using the transfer matrix technique [3], which relates the conductance G with the transmission matrix t, i.e. $G=(e^2/h)g$, with $g=2\text{Tr}(tt^\dagger)$. The g defined here is for both spins. In Fig. 1 we present the results of $p_c(g)$ for three different sizes of N=5, 10, and 20.

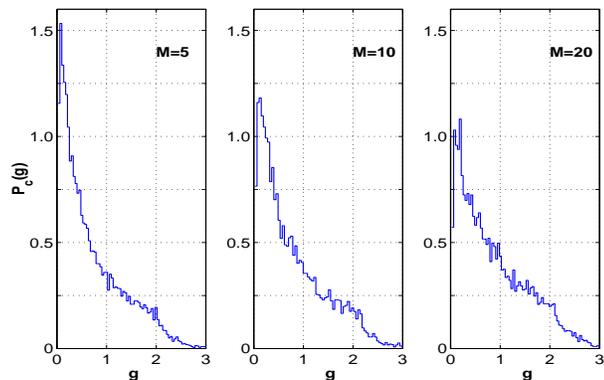

FIG. 1. The distribution of g at the critical point for three different sizes

The mobility edge [1] is at W=16.5 and E=0.0. Notice that as the size of the system increases a dip is developed at g=0, which is not present in the results presented in Fig. 2 of Ref. 1. We therefore have a size dependent $p_c(g)$, which has a dip at small g. It is well known that p(g) for extended states is gaussian, while for localized states is log-normal. However, it is not well known either experimentally [4] or theoretically what is the correct form of the probability distribution at the mobility edge. $p_c(g)$ obtained [2] in the $\epsilon$ expansion in the field theory has a hole at small g in agreement with the numerical results presented here. Recent results [5] for a 2d TB model in the presence of a strong magnetic field show that $p_c(g)$ is very broad with a dip at small g. The 2d $p_c(g)$ is very different from the 3d $p_c(g)$ presented here. We have also calculated the average value of the conductance at the critical point (E=0.0 and W=16.5) for N=5, 10 and 20 for 20000, 10000 and 8000 random configurations respectively. The results are summarized in table I.

TABLE I. The means and standard deviations of the critical distribution of g and ln g.

| N  | $<g>$ | $\sigma_g$ | $<lng>$ | $\sigma_{lng}$ |
|----|-------|------------|---------|----------------|
| 5  | 0.72  | 0.64       | -0.857  | 1.19           |
| 10 | 0.78  | 0.66       | -0.727  | 1.11           |
| 20 | 0.86  | 0.68       | -0.587  | 1.09           |

Notice that both g and $lng$ have very large standard deviations, as big as their average values. Our results for both $<g>$ and $<g>_g = e^{<lng>}$ for the N=10 case (0.78, 0.48) are larger than the results presented (0.58, 0.30) in table III of Ref. 1 for the same model. This difference might be due [6] to the different boundary conditions used by Ref. 1 (fixed) and ourselves (periodic). For the 2d case [5] it is shown that $<g>=1.00$ and $<g>_g=0.88$ for the infinite size system. If we extrapolate our finite-size results to infinite sizes we obtain that $<g>=1.00$ and $<g>_g=0.70$. Remember that $\sigma_g$ is comparable to $<g>$.

In summary, we have numerically calculated the full probability distribution of the conductance, $p_c(g)$, at the Anderson critical point. We find that $p_c(g)$ has a dip at small g in agreement with the $\epsilon$ expansion results [2]. The $p_c(g)$ for the 3d system is quite different from that of the 2d quantum critical point [5]. The universality or not of these distributions is of central importance to the field of disordered systems.

C. M. Soukoulis, Xiaosha Wang, Qiming Li and M. M. Sigalas

Ames Laboratory and Dept. of Physics and Astronomy, Iowa State University, 50011, U.S.A.